\documentstyle[twocolumn,aps,epsf]{revtex}
\begin{document}
\title{Comment on ``Plateaus Observed in the Field profile of Thermal Conductivity
in the Superconductor Bi$_2$Sr$_2$CaCu$_2$O$_8$'' }
\author{H. Aubin, K. Behnia}
\address{Laboratoire de Physique des Solides (CNRS), Universit\'e Paris-Sud, 91405
Orsay, France }
\author{S. Ooi and T. Tamegai}
\address{Department of Applied Physics, The University of Tokyo, 7-3-1 Hongo, Bunkyo-ku Tokyo 113, Japan}
\date{November 10, 1997}
\maketitle

Recently, Krishana {\it et al.}\cite{Krishana} reported that the thermal
conductivity, $\kappa $, of Bi$_2$Sr$_2$CaCu$_2$O$_8$ at low temperature 
becomes field-independent above a temperature-dependent threshold field 
H$_k$(T). This remarkable result indicates a phase transition separating a 
low-field state where the thermal conductivity decreases with increasing field 
and a high-field one where it is insensitive to the applied magnetic field. The authors argue that this phase transition is not related to the vortex lattice 
because of the temperature-dependence of H$_k$(T) (roughly proportional to 
T$^2$) as well as its magnitude. Instead, they suggest a field-induced 
electronic phase transition leading to a sudden vanishing of the quasi-particle 
contribution to the heat transport. One possible scenario would be the
introduction of an additional id$_{xy}$ component to the parent d$_{x^2-y^2}$
superconducting order parameter above the threshold field.

In the mixed state of high-T$_c$ cuprates, the magnitude of $\kappa $ can 
depend on the magneto-thermal history of the sample which
leads to different field profiles\cite{Richardson}. Is the field-independent
thermal conductivity of the high-field state insensentive to the way the
magnetic field is applied? Ref. 1 does not address this question.

In order to gain insight on the nature of this phase transition, we studied
the  thermal conductivity of a Bi$_2$Sr$_2$CaCu$_2$O$_8$ single crystal as a
function of a magnetic field ramped up and down and then reversed. This procedure is similar to the one used in the magnetization studies. Fig.1 shows the results at T = 8.4 K.  On the top panel, beginning with a Zero-Field Cooled (ZFC) 
sample, the thermal conductivity decreases with increasing magnetic field 
and at about  1.5 T ( $\approx $ H$_k$(T= 8.4 K) in Ref.1), a kink occurs in 
$\kappa $(H) followed by a quasi-constant thermal conductivity up to 5 T. Then, the magnetic field is decreased and, surprisingly, a sharp drop of $\kappa$ is 
observed over a small range in magnetic field (0.2 T), followed by a second 
plateau with a lower magnitude. At 2 T, $\kappa $ begins to 
increase again, but it does not attain its initial ZFC magnitude which indicates 
that trapped vortices affect thermal conductivity.
 As seen in the figure, the same sequence of events occur when the measurements are pursued to negative values of magnetic field. The bottom panel represents  
subsequent measurements for fields ramped up and down to 10 T and -1.3 T. The 
same features are present and the magnitude of the drop in $\kappa $ at 10 T is comparable to what was observed at 5 T. Note that the drop occurs concomitantly with 
the sign change in the irreversible magnetization. In a simple Bean model, this is related to a modification of field profile in the sample for ascending 
and descending fields.

%%%%%%%%%%%%%%%%%%%%%%%%%%%%%%%%%%%%%%%%%%%%%%%%%%%%%%%%%%%%%%%%%%%%%%%%%%%%%
\begin{figure}[tbp]
\epsfxsize=8.5cm
$$\epsffile{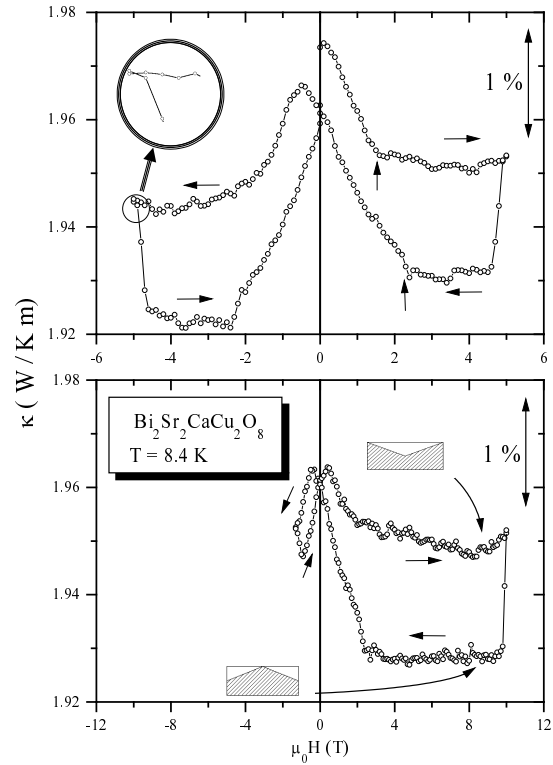}$$
\caption{Field-dependence of thermal conductivity at T=8.4K. The field was
ramped up and down in the directions indicated by arrows. Note the schematic 
field profile in the sample for the two ramping directions (demagnetization 
effects are negletced).}
\label{fig1}
\end{figure}
%%%%%%%%%%%%%%%%%%%%%%%%%%%%%%%%%%%%%%%%%%%%%%%%%%%%%%%%%%%%%%%%%%%%%%%%%%%%%

To explain the insensitivity of thermal conductivity in the plateau regime,
Krishana {\it et al.} invoked two independent constraints. The first one
implies no heat transport by quasi-particles in fields above H$_k$ and the second regards
the absence of vortex scattering of phonons. According to their picture, the
background thermal conductivity is exclusively due to phonons which do not
``see'' the vortices. Our findings show that this background depends on the
field profile in the sample which is incompatible with this picture. In
conclusion, we think that alternative scenarios for this anomaly must be 
considered, including those involving the vortex lattice.

\end{document}